
%
%
%
\hsize=6.5truein
\hoffset=0truein
\vsize=8.9truein
\voffset=0truein
\font\twelverm=cmr10 scaled 1200    \font\twelvei=cmmi10 scaled 1200
\font\twelvesy=cmsy10 scaled 1200   \font\twelveex=cmex10 scaled 1200
\font\twelvebf=cmbx10 scaled 1200   \font\twelvesl=cmsl10 scaled 1200
\font\twelvett=cmtt10 scaled 1200   \font\twelveit=cmti10 scaled 1200
\skewchar\twelvei='177   \skewchar\twelvesy='60
\def\twelvepoint{\normalbaselineskip=12.4pt
  \abovedisplayskip 12.4pt plus 3pt minus 9pt
  \belowdisplayskip 12.4pt plus 3pt minus 9pt
  \abovedisplayshortskip 0pt plus 3pt
  \belowdisplayshortskip 7.2pt plus 3pt minus 4pt
  \smallskipamount=3.6pt plus1.2pt minus1.2pt
  \medskipamount=7.2pt plus2.4pt minus2.4pt
  \bigskipamount=14.4pt plus4.8pt minus4.8pt
  \def\rm{\fam0\twelverm}          \def\it{\fam\itfam\twelveit}%
  \def\sl{\fam\slfam\twelvesl}     \def\bf{\fam\bffam\twelvebf}%
  \def\mit{\fam 1}                 \def\cal{\fam 2}%
  \def\tt{\twelvett}
  \textfont0=\twelverm   \scriptfont0=\tenrm   \scriptscriptfont0=\sevenrm
  \textfont1=\twelvei    \scriptfont1=\teni    \scriptscriptfont1=\seveni
  \textfont2=\twelvesy   \scriptfont2=\tensy   \scriptscriptfont2=\sevensy
  \textfont3=\twelveex   \scriptfont3=\twelveex  \scriptscriptfont3=\twelveex
  \textfont\itfam=\twelveit
  \textfont\slfam=\twelvesl
  \textfont\bffam=\twelvebf \scriptfont\bffam=\tenbf
  \scriptscriptfont\bffam=\sevenbf
  \normalbaselines\rm}

\font\titlerm=cmr10 scaled\magstep3
\font\titleit=cmti10 scaled\magstep3

\def\titlefonts{\def\rm{\titlerm}
                \def\it{\titleit} \rm}
\font\twelvesc=cmcsc10 scaled 1200

\def\beginlinemode{\endmode
  \begingroup\parskip=0pt \obeylines\def\\{\par}\def\endmode{\par\endgroup}}
\def\beginparmode{\endmode
  \begingroup \def\endmode{\par\endgroup}}
\let\endmode=\par
{\obeylines\gdef\
{}}
\def\singlespace{\baselineskip=\normalbaselineskip}
\def\oneandahalfspace{\baselineskip=\normalbaselineskip
  \multiply\baselineskip by 3 \divide\baselineskip by 2}
\def\doublespace{\baselineskip=\normalbaselineskip \multiply\baselineskip by 2}

\newcount\firstpageno
\firstpageno=2
\footline={\ifnum\pageno<\firstpageno{\hfil}\else{\hfil\twelverm\folio\hfil}\fi}
\let\rawfootnote=\footnote		
\def\footnote#1#2{{\rm\singlespace\parindent=0pt\rawfootnote{#1}{#2}}}
\def\raggedcenter{\leftskip=4em plus 12em \rightskip=\leftskip
  \parindent=0pt \parfillskip=0pt \spaceskip=.3333em \xspaceskip=.5em
  \pretolerance=9999 \tolerance=9999
  \hyphenpenalty=9999 \exhyphenpenalty=9999 }
\parskip=\medskipamount
\twelvepoint		
\doublespace		
\overfullrule=0pt	

\def\preprintno#1{
 \rightline{\rm #1}}	

\def\title                      
  {\null\vskip 3pt plus 0.2fill
   \beginlinemode \doublespace \raggedcenter \titlefonts}

\def\author                     
  {\vskip 3pt plus 0.2fill \beginlinemode
   \singlespace \raggedcenter\twelvesc}

\def\affil			
  {\vskip 3pt plus 0.03fill \beginlinemode
   \oneandahalfspace \it \centerline}

\def\abstract			
  {\vskip 3pt plus 0.3fill \beginparmode
   \doublespace \narrower ABSTRACT: }

\def\endtitlepage		
  {\endpage			
   \body}

\def\body			
  {\beginparmode}		

\def\head#1{			
  \filbreak\vskip 0.5truein	
  {\immediate\write16{#1}
   \raggedcenter \bf\uppercase{#1}\par}
   \nobreak\vskip 0.25truein\nobreak}

\def\subhead#1{			
  \vskip 0.25truein		
  {\raggedcenter \bf{#1}\par}
   \nobreak\vskip 0.25truein\nobreak}

\def\itsubhead#1{		
  \vskip 0.25truein		
  {\noindent\it{#1}\par}
   \nobreak\vskip 0.2truein\nobreak}

\def\references
  {\head{REFERENCES}
   \frenchspacing \parindent=0pt \leftskip=0.8truecm \rightskip=0truecm
   \parskip=4pt plus 2pt \everypar{\hangindent=\parindent}}

\def\pr{\journal Phys. Rev.}

\def\rmp{\journal Rev. Mod. Phys.}

\def\np{\journal Nucl. Phys.}
\def\pl{\journal Phys. Lett.}

\def\anphys{\journal Ann. Phys.}
\def\zphys{\journal Z. Phys.}

\def\jps{\journal J. Phys. (Sov.)}
\def\ptp{\journal Prog. Theor. Phys.}

\def\endreferences{\body}

\def\figurecaptions		
  {\endpage
   \beginparmode
   \head{Figure Captions}
}

\def\endpage			
  {\vfill\eject}

\def\endpaper			
  {\endmode\vfill\supereject}
\def\endit
  {\endpaper\end}

\def\frac#1#2{{\textstyle{#1 \over #2}}}
\def\ref#1{ref. #1}
\def\Ref#1{Ref. #1}

\def\eg{{\it e.g.}}

\def\ie{{\it i.e.}}

\def\etc{{\it etc.}}

\def\sla{\raise.15ex\hbox{$/$}\kern-.57em}
\def\leaderfill{\leaders\hbox to 1em{\hss.\hss}\hfill}
\def\twiddle{\lower.9ex\rlap{$\kern-.1em\scriptstyle\sim$}}
\def\bigtwiddle{\lower1.ex\rlap{$\sim$}}
\def\gtwid{\mathrel{\raise.3ex\hbox{$>$\kern-.75em\lower1ex\hbox{$\sim$}}}}
\def\ltwid{\mathrel{\raise.3ex\hbox{$<$\kern-.75em\lower1ex\hbox{$\sim$}}}}
\def\square{\kern1pt\vbox{\hrule height 1.2pt\hbox{\vrule width 1.2pt\hskip 3pt
   \vbox{\vskip 6pt}\hskip 3pt\vrule width 0.6pt}\hrule height 0.6pt}\kern1pt}

%
%
%
\def\refstylenp{		
  \gdef\refto##1{~[##1]}				
  \gdef\r##1{~[##1]}	         			
  \gdef\refis##1{\indent\hbox to 0pt{\hss[##1]~}}     	
  \gdef\citerange##1##2##3{\cite{##1}--\setbox0=\hbox{\cite{##2}}\cite{##3}}
  \gdef\rrange##1##2##3{~[\cite{##1}--\setbox0=\hbox{\cite{##2}}\cite{##3}]}
  \gdef\journal##1, ##2, ##3,                           
    ##4,{{\sl##1} {\bf ##2} (##3) ##4}}
\def\refstylezphys{		
  \gdef\refto##1{~[##1]}				
  \gdef\r##1{~[##1]}	         			
  \gdef\refis##1{\indent\hbox to 0pt{\hss[##1]~}}     	
  \gdef\citerange##1##2##3{\cite{##1}--\setbox0=\hbox{\cite{##2}}\cite{##3}}
  \gdef\rrange##1##2##3{~[\cite{##1}--\setbox0=\hbox{\cite{##2}}\cite{##3}]}
  \gdef\journal##1, ##2, ##3,                           
    ##4,{{\sl##1} {\bf ##2}, ##4 (##3)}}
\def\refstylepr{		
  \gdef\refto##1{~[##1]}		
  \gdef\r##1{~[##1]}		        
  \gdef\refis##1{\indent\hbox to 0pt{\hss[##1]~}}	
  \gdef\citerange##1##2##3{\cite{##1}--\setbox0=\hbox{\cite{##2}}\cite{##3}}
  \gdef\rrange##1##2##3{~[\cite{##1}--\setbox0=\hbox{\cite{##2}}\cite{##3}]}
  \gdef\journal##1, ##2, ##3,                           
    ##4,{{\sl##1} {\bf ##2}, ##4 (##3)}}
\def\refstyleijmp{		
  \gdef\refto##1{$^{##1}$}				
  \gdef\r##1{$^{##1}$}	         			
  \gdef\refis##1{\indent\hbox to 0pt{\hss##1.~}}     	
  \gdef\citerange##1##2##3{\cite{##1}--\setbox0=\hbox{\cite{##2}}\cite{##3}}
  \gdef\rrange##1##2##3{$^{\cite{##1}-\setbox0=\hbox{\cite{##2}}\cite{##3}}$}
  \gdef\journal##1, ##2, ##3,                           
    ##4,{{\sl##1} {\bf ##2} (##3) ##4}}
\def\(#1){(\call{#1})}
\def\call#1{{#1}}

\def\osu
{Department of Physics, The Ohio State University, Columbus, OH 43210, USA}
\def\mpihd
{Max-Planck-Institut f\"ur Kernphysik, D-69029 Heidelberg 1, Germany}

\def\ths{\thinspace}
\def\a{\alpha}
\def\b{\beta}

\def\e{\epsilon}
\def\g{\gamma}
\def\d{\delta}
\def\p{\phi}

\def\l{\lambda}
\def\L{\Lambda}

\def\vp{\varphi}
\def\del{\partial}
\def\ha{{1\over 2}}
\def\exp{\hbox{exp}}

\def\ra{\rightarrow}
\def\bar#1{\overline{ #1 }}
\def\psibar{\overline{\psi}}
\def\slash#1{/\!\!\!\!{#1}}
\def\pri{\prime}

\def\lra{\leftrightarrow}

\def\ub#1{\underline{#1}}
\def\senk#1{ {\vec #1}_\perp }
\def\kd3{\delta^{(3)}}
\def\sym#1,#2{\Biggl\{ {#1}{1\over i\del^+} {#2} \Biggr\}_{\rm sym} }
\def\ssym#1,#2{\Biggl\{ {#1}{1\over (i\del^+)^2} {#2} \Biggr\}_{\rm sym} }
\refstylenp
\catcode`@=11
\newcount\tagnumber\tagnumber=0
\immediate\newwrite\eqnfile
\newif\if@qnfile\@qnfilefalse
\def\write@qn#1{}
\def\writenew@qn#1{}
\def\w@rnwrite#1{\write@qn{#1}\message{#1}}
\def\@rrwrite#1{\write@qn{#1}\errmessage{#1}}
\def\taghead#1{\gdef\t@ghead{#1}\global\tagnumber=0}
\def\t@ghead{}
\expandafter\def\csname @qnnum-3\endcsname
  {{\t@ghead\advance\tagnumber by -3\relax\number\tagnumber}}
\expandafter\def\csname @qnnum-2\endcsname
  {{\t@ghead\advance\tagnumber by -2\relax\number\tagnumber}}
\expandafter\def\csname @qnnum-1\endcsname
  {{\t@ghead\advance\tagnumber by -1\relax\number\tagnumber}}
\expandafter\def\csname @qnnum0\endcsname
  {\t@ghead\number\tagnumber}
\expandafter\def\csname @qnnum+1\endcsname
  {{\t@ghead\advance\tagnumber by 1\relax\number\tagnumber}}
\expandafter\def\csname @qnnum+2\endcsname
  {{\t@ghead\advance\tagnumber by 2\relax\number\tagnumber}}
\expandafter\def\csname @qnnum+3\endcsname
  {{\t@ghead\advance\tagnumber by 3\relax\number\tagnumber}}
\def\equationfile{%
  \@qnfiletrue\immediate\openout\eqnfile=\jobname.eqn%
  \def\write@qn##1{\if@qnfile\immediate\write\eqnfile{##1}\fi}
  \def\writenew@qn##1{\if@qnfile\immediate\write\eqnfile
    {\noexpand\tag{##1} = (\t@ghead\number\tagnumber)}\fi}
}
\def\callall#1{\xdef#1##1{#1{\noexpand\call{##1}}}}
\def\call#1{\each@rg\callr@nge{#1}}
\def\each@rg#1#2{{\let\thecsname=#1\expandafter\first@rg#2,\end,}}
\def\first@rg#1,{\thecsname{#1}\apply@rg}
\def\apply@rg#1,{\ifx\end#1\let\next=\relax%
\else,\thecsname{#1}\let\next=\apply@rg\fi\next}
\def\callr@nge#1{\calldor@nge#1-\end-}
\def\callr@ngeat#1\end-{#1}
\def\calldor@nge#1-#2-{\ifx\end#2\@qneatspace#1 %
  \else\calll@@p{#1}{#2}\callr@ngeat\fi}
\def\calll@@p#1#2{\ifnum#1>#2{\@rrwrite{Equation range #1-#2\space is bad.}
\errhelp{If you call a series of equations by the notation M-N, then M and
N must be integers, and N must be greater than or equal to M.}}\else%
 {\count0=#1\count1=#2\advance\count1
by1\relax\expandafter\@qncall\the\count0,%
  \loop\advance\count0 by1\relax%
    \ifnum\count0<\count1,\expandafter\@qncall\the\count0,%
  \repeat}\fi}
\def\@qneatspace#1#2 {\@qncall#1#2,}
\def\@qncall#1,{\ifunc@lled{#1}{\def\next{#1}\ifx\next\empty\else
  \w@rnwrite{Equation number \noexpand\(>>#1<<) has not been defined yet.}
  >>#1<<\fi}\else\csname @qnnum#1\endcsname\fi}
\let\eqnono=\eqno
\def\eqno(#1){\tag#1}
\def\tag#1$${\eqnono(\displayt@g#1 )$$}
\def\aligntag#1\endaligntag
  $${\gdef\tag##1\\{&(##1 )\cr}\eqalignno{#1\\}$$
  \gdef\tag##1$${\eqnono(\displayt@g##1 )$$}}

\def\eqalignno#1{\displ@y \tabskip\centering
  \halign to\displaywidth{\hfil$\displaystyle{##}$\tabskip\z@skip
    &$\displaystyle{{}##}$\hfil\tabskip\centering
    &\llap{$\displayt@gpar##$}\tabskip\z@skip\crcr
    #1\crcr}}
\def\displayt@gpar(#1){(\displayt@g#1 )}
\def\displayt@g#1 {\rm\ifunc@lled{#1}\global\advance\tagnumber by1
        {\def\next{#1}\ifx\next\empty\else\expandafter
        \xdef\csname @qnnum#1\endcsname{\t@ghead\number\tagnumber}\fi}%
  \writenew@qn{#1}\t@ghead\number\tagnumber\else
        {\edef\next{\t@ghead\number\tagnumber}%
        \expandafter\ifx\csname @qnnum#1\endcsname\next\else
        \w@rnwrite{Equation \noexpand\tag{#1} is a duplicate number.}\fi}%
  \csname @qnnum#1\endcsname\fi}
\def\ifunc@lled#1{\expandafter\ifx\csname @qnnum#1\endcsname\relax}
\let\@qnend=\end\gdef\end{\if@qnfile
\immediate\write16{Equation numbers written on []\jobname.EQN.}\fi\@qnend}
\newcount\r@fcount \r@fcount=0
\newcount\r@fcurr
\immediate\newwrite\reffile
\newif\ifr@ffile\r@ffilefalse
\def\w@rnwrite#1{\ifr@ffile\immediate\write\reffile{#1}\fi\message{#1}}
\def\writer@f#1>>{}
\def\referencefile{
  \r@ffiletrue\immediate\openout\reffile=\jobname.ref%
  \def\writer@f##1>>{\ifr@ffile\immediate\write\reffile%
    {\noexpand\refis{##1} = \csname r@fnum##1\endcsname = %
     \expandafter\expandafter\expandafter\strip@t\expandafter%
     \meaning\csname r@ftext\csname r@fnum##1\endcsname\endcsname}\fi}%
  \def\strip@t##1>>{}}

\def\citeall#1{\xdef#1##1{#1{\noexpand\cite{##1}}}}
\def\cite#1{\each@rg\citer@nge{#1}}	
\def\each@rg#1#2{{\let\thecsname=#1\expandafter\first@rg#2,\end,}}
\def\first@rg#1,{\thecsname{#1}\apply@rg}	
\def\apply@rg#1,{\ifx\end#1\let\next=\relax
\else,\thecsname{#1}\let\next=\apply@rg\fi\next}
\def\citer@nge#1{\citedor@nge#1-\end-}	
\def\citer@ngeat#1\end-{#1}
\def\citedor@nge#1-#2-{\ifx\end#2\r@featspace#1 
  \else\citel@@p{#1}{#2}\citer@ngeat\fi}	
\def\citel@@p#1#2{\ifnum#1>#2{\errmessage{Reference range #1-#2\space is bad.}%
    \errhelp{If you cite a series of references by the notation M-N, then M and
    N must be integers, and N must be greater than or equal to M.}}\else%
 {\count0=#1\count1=#2\advance\count1
by1\relax\expandafter\r@fcite\the\count0,%
  \loop\advance\count0 by1\relax
    \ifnum\count0<\count1,\expandafter\r@fcite\the\count0,%
  \repeat}\fi}
\def\r@featspace#1#2 {\r@fcite#1#2,}	
\def\r@fcite#1,{\ifuncit@d{#1}
    \newr@f{#1}%
    \expandafter\gdef\csname r@ftext\number\r@fcount\endcsname%
                     {\message{Reference #1 to be supplied.}%
                      \writer@f#1>>#1 to be supplied.\par}%
 \fi%
 \csname r@fnum#1\endcsname}
\def\ifuncit@d#1{\expandafter\ifx\csname r@fnum#1\endcsname\relax}%
\def\newr@f#1{\global\advance\r@fcount by1%
    \expandafter\xdef\csname r@fnum#1\endcsname{\number\r@fcount}}
\let\r@fis=\refis			
\def\refis#1#2#3\par{\ifuncit@d{#1}
   \newr@f{#1}%
   \w@rnwrite{Reference #1=\number\r@fcount\space is not cited up to now.}\fi%
  \expandafter\gdef\csname r@ftext\csname r@fnum#1\endcsname\endcsname%
  {\writer@f#1>>#2#3\par}}
\def\ignoreuncited{
   \def\refis##1##2##3\par{\ifuncit@d{##1}%
     \else\expandafter\gdef\csname r@ftext\csname
r@fnum##1\endcsname\endcsname%
     {\writer@f##1>>##2##3\par}\fi}}
\def\r@ferr{\endreferences\errmessage{I was expecting to see
\noexpand\endreferences before now;  I have inserted it here.}}
\let\r@ferences=\references
\def\references{\r@ferences\def\endmode{\r@ferr\par\endgroup}}
\let\endr@ferences=\endreferences
\def\endreferences{\r@fcurr=0
  {\loop\ifnum\r@fcurr<\r@fcount
    \advance\r@fcurr by 1\relax\expandafter\r@fis\expandafter{\number\r@fcurr}%
    \csname r@ftext\number\r@fcurr\endcsname%
  \repeat}\gdef\r@ferr{}\endr@ferences}
\let\r@fend=\endpaper\gdef\endpaper{\ifr@ffile
\immediate\write16{Cross References written on []\jobname.REF.}\fi\r@fend}
\catcode`@=12
\citeall\refto		
\citeall\ref		%
\citeall\Ref		%
\citeall\r		%
\ignoreuncited

\def\uk{\ub{k}}
\def\up{\ub{p}}
\def\uq{\ub{q}}
\def\ukp{\ub{k}^\prime}
\def\upp{\ub{p}^\prime}
\def\senk#1{{#1}_\perp}

\def\szm#1{\langle #1 \rangle_{o}}
\def\nm#1{#1_n}
\def\totzm#1{\langle #1 \rangle }

\def\zm#1{{\buildrel{\circ} \over {#1}}{} }
\def\ie{{\it i.e.}}
\def\jcp{\journal J. Comp. Phys.}

\singlespace
\preprintno{MPIH--V10--1994}
\preprintno{OSU--NT--94--03}
\preprintno{May 1994}
\doublespace

\title On the Discretized Light-Cone Quantization of Electrodynamics
\author Alex C. Kalloniatis$^a$ and David G. Robertson$^b$

\affil{$^a$\mpihd}
\affil{$^b$\osu}

\abstract\
Discretized light-cone quantization of (3+1)-dimensional
electrodynamics is discussed, with careful attention paid to the
interplay between gauge choice and boundary conditions.  In the zero
longitudinal momentum sector of the theory a general gauge fixing is
performed, and the corresponding relations that determine the zero
modes of the gauge field are obtained.  One particularly natural gauge
choice in the zero mode sector is identified, for which the constraint
relations are simplest and the fields may be taken to satisfy the
usual canonical commutation relations.  The constraints are solved in
perturbation theory, and the Poincar\'e generators $P^\mu$ are
constructed.  The effect of the zero mode contributions on the
one-loop fermion self-energy is studied.


\endtitlepage
\oneandahalfspace

\subhead{1. Introduction}
\taghead{1.}

Light-cone quantization, or more properly quantization on a null
plane\r{dirac49}, seems to offer several advantages over the more
traditional equal-time quantization for a nonperturbative treatment of
field theories.  There are arguments, for example, that certain
Lorentz boosts are in the kinematical subgroup, and that the vacuum
structure is simpler.  There has recently been considerable effort
devoted to exploiting these advantages in the context of a
Tamm-Dancoff-style solution of field theory\r{tamm45,dancoff50}.  For
an overview of this work with many references see ref.\r{bmpp93}.
This approach has been strikingly successful in two spacetime
dimensions\rrange{pb85}{epb87,hv87,hv88,burkardt89,hpb90,hps91}{mp93},
and encouraging results have also been obtained recently in
four-dimensional models\rrange{tbp91}{kpw92,kp92}{ghpsw93}.  For a
complete attack on four-dimensional QCD to be feasible, however, many
technical obstacles remain to be overcome.

A particularly simple framework for actual calculations is that of
``discretized'' light-cone quantization (DLCQ), in which the theory is
defined on a light-cone ``torus''\r{pb85}.  It then possesses a
discrete momentum-space basis, which regulates infrared divergences
and is ready-made for numerical analysis.  The goal of this approach
is to give a controllable formulation of a quantum field theory in
terms of a Hamiltonian eigenvalue equation, and then to solve it, in
general numerically.

Certain formal aspects of this approach, however, are not yet
completely under control.  One such area concerns the zero modes (in
the Fourier sense) of bosonic fields.  To illustrate the basic problem
let us consider a self-interacting scalar field in $1+1$ dimensions,
for which the Euler-Lagrange equation is
$$
(4 \partial_- \partial_+  + m^2) \phi = - \lambda \phi^3\; .
					\eqno(scalareom)
$$
(Our notational conventions are summarized in Appendix A.)  After
imposing periodic boundary conditions on the finite interval $-L\leq
x^-\leq L$ we can expand the field in discrete Fourier modes with
momenta $ k^+ = {{n \pi} \over {L}},$ $n$ an even integer.  We then
project out the zero mode ($n=0$) by integrating both sides of the
equation over the entire interval, obtaining
$$
m^2 \zm{\phi} = -\lambda \zm{\phi}^3 - {\lambda\over2L}
\int_{-L}^{L} dx^-
( 3 \zm{\phi} \varphi^2 + \varphi^3 ) \; .
					\eqno(phicons)
$$
Here $\zm{\phi}$ is the zero mode and $\varphi$ is the complementary
``normal mode'' part: $\varphi\equiv\phi-\zm{\phi}$.  The important
thing to note is that the time ($x^+$) derivative has dropped out due
to the chosen boundary conditions.  The zero mode is therefore a {\it
constrained} field for which we cannot specify independent quantum
commutation relations\rrange{my76}{wittman88,hkw91b}{mr92}.
Furthermore, $\phi_0$ is needed for the computation of \eg\ the
Poincar\'e generators.  The nonlinear operator constraint \(phicons)
must therefore be solved before the Hamiltonian can even be written
down.

There is a striking simplification that occurs elsewhere in the
theory, however: the Fock vacuum is an exact eigenstate of the full
Hamiltonian.  This follows from light-cone momentum $(P^+)$
conservation and the observation that the zero mode does not
correspond to a degree of freedom---that is, there is no $P^+=0$
quantum in the theory.  The bare vacuum is thus the only state in the
theory with $P^+=0$, and must therefore be an exact eigenstate of the
full Hamiltonian.  This is a highly desirable feature if we wish to
have a constituent picture of relativistic bound states, and describe,
for example, a baryon as primarily a three-quark state plus a few
higher Fock states {\it \'a la} Tamm and Dancoff.  In equal-time
quantization, where the physical vacuum state is an infinite
superposition of states with arbitrarily large numbers of bare quanta,
it would be extremely difficult to describe a baryon in this fashion.
In this case a sensible constituent description would be in terms of
``quasiparticles,'' perhaps corresponding loosely to the quarks of the
constituent quark model.  These would be complicated collective
excitations above the physical vacuum state.  The difficulty here, of
course, is that without knowledge of the full solution of the theory
we have no idea how to connect these quasiparticle states to the bare
states (in terms of which the Hamiltonian is easily formulated).

In DLCQ, the problem of the vacuum is apparently shifted to that of
obtaining solutions to the constraint equations for the zero modes.
Some preliminary support for this view is provided by considering the
model of eq. \(scalareom) with $m^2<0,$ in which case we anticipate
spontaneous breakdown of the reflection symmetry $\p\ra-\p$.  Here we
find\r{hksw92,dgr93,bpv93} that there are multiple solutions to the
constraint \(phicons), and we must choose one to use in formulating
the theory.  This choice is analogous to what in the conventional
language we would call the choice of vacuum state.  The various
solutions contain $c$-number pieces which produce the possible vacuum
expectation values of $\p.$ The properties of the strong-coupling
phase transition in this model are also reproduced, including its
second-order nature and a reasonable value for the critical
coupling\r{bpv93,pv94}.  (For an earlier study of this phase
transition in DLCQ without the zero mode see\r{hv88b}.)  So in some
cases, at least, physics which we normally associate with the vacuum
can be manifested in these zero modes, in a formalism where the vacuum
state itself is simple.

We should perhaps emphasize that, apart from the question of whether
or not VEVs arise, solving the constraint equations really amounts to
determining the Hamiltonian (and other Poincar\'e generators).  In
general, $P^-$ becomes very complicated when the zero mode
contributions are included; this is in some sense the price we pay to
achieve a formulation with a simple vacuum.$^{\ddagger1}$
\footnote{}{$^{\ddagger1}$ It may be possible to think of the
discretization as a cutoff which removes states with $0<p^+<\pi/L,$
and the zero mode contributions to the Hamiltonian as effective
interactions that restore the discarded physics. We shall not pursue
this idea in detail here, except to note that from the light-cone
power counting analysis of Wilson\r{wilsonunpub,wwhzpg94} it is clear
that there will be a huge number of allowed operators.}
The other Poincar\'e generators apparently also receive contributions
from the zero modes, and it becomes important to check whether the
supposedly kinematical ones---$P^+$ and $P^i$---remain kinematical
when the zero modes are included.  As we shall see, this is not always
guaranteed to be the case.

When considering a gauge theory there is another ``zero mode'' problem
associated with the choice of gauge in the compactified case.  This
subtlety, however, is not particular to the light cone; indeed, its
occurrence is quite familiar in equal-time quantization on a
torus\r{gaugerefs}.  In the present context, the difficulty is that
the zero mode in $A^+$ is in fact gauge-invariant, so that the
light-cone gauge $A^+=0$ cannot be reached.  Thus we have a pair of
interconnected problems: first, a practical choice of gauge; and
second, the presence of constrained zero modes of the gauge field.  In
two recent papers\r{kp93,kp94} these two problems were separated and
consistent gauge fixing conditions were introduced to allow isolation
of the dynamical and constrained fields.  In the present paper we
shall generalize the gauge fixing described in ref.\r{kp94}, and
construct the Poincar\'e generators in perturbation theory.  Our aim
is to study the types of operators induced by the zero modes, and to
examine their effects on the perturbative renormalization of this
theory.

We begin in the next section by reviewing the approach laid out in
ref.\r{kp94}, and separating the constrained zero modes from the
dynamical degrees of freedom within a general gauge fixing.  We obtain
the relations that define the constrained operators, and show that
there is a unique choice of gauge in the zero mode sector for which
the naively kinematical operators $P^i$ retain their free-field forms,
so that the usual (free-field) commutation relations can be employed.
In section 3 we solve the constraints in perturbation theory, and
construct the Hamiltonian to lowest nontrivial order.  In section 4 we
study the effects of the zero mode contributions on the one-loop
fermion self-energy.  Some discussion and our conclusions are
presented in section 5.

\vfill\eject

\subhead{2. Gauge Fixing and the Zero-Momentum Modes}
\taghead{2.}

With the standard Lagrangian density for electrodynamics
$$
{\cal L} = - {1\over4} F^{\mu \nu} F_{\mu \nu} +
\bar{\psi} (i \slash{D} - m) \psi
					\eqno(lagrangian)
$$
the equations of motion are the familiar Dirac equation
$$
(i\slash{D}-m)\psi = 0\; ,
					\eqno(dirac)
$$
and Maxwell's equations
$$
\del_\mu F^{\mu\nu} = gJ^\nu\; ,
					\eqno(maxwell)
$$
where $D_\mu=\del_\mu+igA_\mu$ and $J^\mu\equiv
:{\psibar\g^\mu\psi}:$.  Written out explicitly in terms of the
various gauge field components and the spinor projections defined in
Appendix A these become
$$
(2i\del_+-gA^-)\psi_+=\Bigl[-i\a^i\del_i+m\b +g\a^i A_i\Bigr]\psi_-
					\eqno(psipeom)
$$
$$
(2i\del_--gA^+)\psi_-=\Bigl[-i\a^i\del_i+m\b +g\a^i A_i\Bigr]\psi_+
					\eqno(psimeom)
$$
$$
2\del_+\del_-A^+ - 2(\del_-)^2 A^- - 2\del_-\del_j A^j
-\del_\perp^2 A^+ = gJ^+
					\eqno(gauss)
$$
$$
2\del_+\del_-A^- - 2(\del_+)^2 A^+ - 2\del_+\del_j A^j
-\del_\perp^2 A^- = gJ^-
					\eqno(ampere)
$$
$$
(4\del_+\del_--\del_\perp^2)A^i + \del_i\del_+A^+
+\del_i\del_-A^- + \del_i\del_jA^j = gJ^i\; .
					\eqno(aperpeom)
$$
Observe that in the traditional treatment, choosing the light-cone
gauge $A^+=0$ enables eq. \(gauss) to be solved for $A^-$. In any
case, the spinor projection $\psi_-$ is constrained and determined by
eq. \(psimeom).

Discretization is achieved by putting the theory in a light-cone
``box,'' with $-L_\perp\leq x^i\leq L_\perp$ and $-L\leq x^-\leq L,$
and imposing boundary conditions on the fields.  The choices of
boundary conditions are constrained by the need to be consistent with
the equations of motion.  Because the gauge field couples to a fermion
bilinear, which is necessarily periodic in all coordinates, $A_\mu$
must be taken to be periodic in both $x^-$ and $x_\perp$.  We have
more flexibility with the fermi field, and it is most convenient to
choose this to be periodic in $x_\perp$ and antiperiodic in $x^-.$
This eliminates the zero longitudinal momentum mode while still
allowing an expansion of the field in a complete set of basis
functions.

The functions used to expand the fields may be taken to be plane
waves, and for periodic fields these will of course include
zero-momentum modes.  Let us define, for a periodic quantity $f$, its
longitudinal zero mode
$$
\szm{f} \equiv {1\over2L}\int_{-L}^L dx^- f(x^-)
					\eqno(zeromode)
$$
and the corresponding normal mode part
$$
\nm{f} \equiv f - \szm{f}\; .
					\eqno(normal)
$$
We shall further denote the ``global zero mode''---the mode
independent of all the spatial coordinates---by $\totzm{f}$:
$$
\totzm{f} \equiv {1\over\Omega} \int_{-L}^{L} dx^-
\int_{-L_\perp}^{L_\perp}d^2x_\perp\ths f(x^-)\; .
					\eqno(ozm)
$$
Finally, the quantity which will be of most interest to us is the
``proper zero mode,'' defined by
$$
\zm{f} \equiv \szm{f} - \totzm{f}\; .
					\eqno(pzmdef)
$$

By integrating over the appropriate direction(s) of space, we can
project the equations of motion onto the various sectors.  Previous
work on the formulation of QED in DLCQ\r{tbp91,tangphd} has
been implicitly carried out in the normal mode sector, and many of
these results may be carried over without modification.  The global
zero mode sector requires some special treatment, and in fact turns
out to be irrelevant for the perturbative calculations we shall
present here.  A brief description of its features and a proof that it
can be ignored to lowest nontrivial order in perturbation theory are
given in Appendix B.

Thus we concentrate our attention on the proper zero mode sector, in
which the equations of motion become
$$
-\del_\perp^2\zm{A}^+ = g\zm{J}^+
					\eqno(zm2)
$$
$$
-2(\del_+)^2\zm{A}^+-\del_\perp^2\zm{A}^--2\del_i\del_+\zm{A}^i = g\zm{J}^-
					\eqno(zm3)
$$
$$
-\del_\perp^2\zm{A}^i+\del_i\del_+\zm{A}^++\del_i\del_j\zm{A}^j = g\zm{J}^i
\; .
					\eqno(zm1)
$$
We first observe that eq. \(zm2), the projection of Gauss' law, is a
constraint which determines the proper zero mode of $A^+$ in terms of
the current $J^+$:

$$
\zm{A}^+ = -g{1\over\del_\perp^2}\zm{J}^+\; .
					\eqno(zm2solved)
$$
(Note that the integral operator $(\del_\perp^2)^{-1}$ is well-defined
in this sector\r{kp94}.)  Eqs. \(zm3) and \(zm1) then determine the
zero modes $\zm{A}^-$ and $\zm{A}^i$.

Eq. \(zm2solved) is clearly incompatible with the strict light-cone
gauge $A^+=0,$ which is most natural in light-cone analyses of gauge
theories.  Here we encounter a common problem in treating axial gauges
on compact spaces\r{gaugerefs}, which has nothing to do with
light-cone quantization {\it per se.} The point is that any
$x^-$-independent part of $A^+$ is in fact gauge-invariant, since
under a gauge transformation
$$
A^+\ra A^+ + 2\del_-\Lambda\; ,
					\eqno(gt)
$$
where $\Lambda$ is a function periodic in all coordinates.$^{\ddagger2}$
\footnote{}{$^{\ddagger2}$ The gauge transformation must also
preserve the boundary conditions on the other fields; thus
\eg\ $\Lambda\sim x^-$ is not in general allowed.  See however
Appendix B.}
Thus it is not possible to bring an arbitrary gauge field
configuration to one satisfying $A^+=0$ via a gauge transformation,
and the light-cone gauge is incompatible with the chosen boundary
conditions.  The closest we can come is to set the normal mode part of
$A^+$ to zero, which is equivalent to
$$
\del_-A^+=0\; .
					\eqno(gauge1)
$$
This condition does not, however, completely fix the gauge---we are
free to make arbitrary $x^-$-independent gauge transformations without
undoing eq. \(gauge1).  We may therefore impose further conditions
on $A_\mu$ in the zero mode sector of the theory.

To see what might be useful in this regard, let us consider solving
eq. \(zm1).  We begin by acting on eq. \(zm1) with $\partial_i$.  The
transverse field $\zm{A}^i$ then drops out and we obtain an expression
for the time derivative of $\zm{A}^+$:
$$
\del_+\zm{A}^+ = g{1\over\del_\perp^2}\del_i \zm{J}^i\; .
					\eqno(dpap)
$$
(This can also be obtained by taking a time derivative of eq.
\(zm2solved), and using current conservation to re-express the right
hand side in terms of $J^i.$)  Inserting this back into eq. \(zm1) we
then find, after some rearrangement,
$$
-\del_\perp^2\Bigl(\d^i_j-{\del_i\del_j\over\del_\perp^2}\Bigr)\zm{A}^j =
g\Bigl(\d^i_j-{\del_i\del_j\over\del_\perp^2}\Bigr)\zm{J}^j\; .
\eqno(Atrans)
$$
Now, the operator $(\d^i_j- \del_i\del_j / \del_\perp^2) $ is nothing
more than the projector of the two-dimensional transverse part of the
vector fields $\zm{A}^i$ and $\zm{J}^i$.  No trace remains of the
longitudinal projection of the field $(\del_i\del_j /
\del_\perp^2)\zm{A}^j$ in eq. \(Atrans).  This reflects precisely the
residual gauge freedom with respect to $x^-$-independent
transformations. To determine the longitudinal part an additional
condition is required.

More concretely, the general solution to eq. \(Atrans) is
$$
\zm{A}^i = -g{1\over\del_\perp^2}\zm{J}^i + \del_i\varphi(x^+,x_\perp)\; ,
					\eqno(secsoln)
$$
where $\varphi$ must be independent of $x^-$ but is otherwise
arbitrary.  Imposing a condition on, say, $\del_i\zm{A}^i$ will
uniquely determine $\vp.$$^{\ddagger3}$
\footnote{}{$^{\ddagger3}$ We could also refuse to completely fix the gauge,
and treat $\varphi$ as a degree of freedom.  It would be unphysical,
however, and would have to be removed by restricting to a suitable
physical subspace.  See ref.\r{mr94} for an example of this type of
construction, in a continuum formulation.}
In ref.\r{kp94}, for example, the condition
$$
\del_i\zm{A}^i=0
					\eqno(compg)
$$
was proposed as being particularly natural. This choice, taken with
the other gauge conditions we have imposed, has been called the
``compactification gauge.''  In this case
$$
\varphi = g{1\over(\del_\perp^2)^2}\del_i \zm{J}^i\; .
					\eqno(compgp)
$$
Of course, other choices are also possible.  For example, we might
generalize eq. \(compgp) to
$$
\vp = \a g{1\over(\del_\perp^2)^2}\del_i\zm{J}^i\; ,
					\eqno(gencompgp)
$$
with $\a$ a real parameter.  The gauge condition corresponding to this
solution is
$$
\del_i \zm{A}^i=-g(1-\a){1\over\del_\perp^2}\del_i \zm{J}^i\; .
					\eqno(zmgaugecond)
$$
We shall refer to this as the ``generalized compactification gauge.''
An arbitrary gauge field configuration $B^\mu $ can be brought to one
satisfying eq. \(zmgaugecond) via the gauge function
$$
\Lambda(x_\perp) =- {1\over\del_\perp^2}
\Bigl[ g (1 - \a) {1\over\del_\perp^2} \del_i \zm{J}^i
+ \del_i \zm{B}^i \Bigr]\; .
					\eqno(gaugetrans)
$$
This is somewhat unusual in that $\Lambda(x_\perp)$ involves the
sources as well as the initial field configuration, but this is
perfectly acceptable.  More generally, $\varphi$ can be any
(dimensionless) function of gauge-invariants constructed from the
fields in the theory, including the currents $J^\pm$. For our purposes
eq. \(zmgaugecond) suffices.

We now have relations defining the proper zero modes of $A^i$,
$$
\zm{A}^i = -g{1\over\del_\perp^2}\Bigl(
\d^i_j-\a{\del_i\del_j\over\del_\perp^2}\Bigr) \zm{J}^j\; ,
					\eqno(aizeromode)
$$
as well as $\zm{A}^+$ (eq. \(zm2solved)).  All that remains is to use the
final constraint, eq. \(zm3), to determine $\zm{A}^-.$  Using eqs.
\(dpap) and \(zmgaugecond), we find that eq. \(zm3) can be written as
$$
\del_\perp^2\zm{A}^- = -g\zm{J}^--2\a g{1\over\del_\perp^2}
\del_+\del_i \zm{J}^i\; .
					\eqno(solvingam0)
$$
After using the equations of motion to express $\del_+\zm{J}^i$ in
terms of the dynamical fields at $x^+=0$, this may be
straightforwardly solved for $\zm{A}^-$ by inverting the
$\del_\perp^2.$ In what follows, however, we shall have no need of
$\zm{A}^-.$ It does not enter the Hamiltonian, for example; as usual,
it plays the role of a multiplier to Gauss' law, eq. \(zm1), which we
are able to implement as an operator identity.

Now, since different choices for $\vp$ merely correspond to different
gauge choices in the zero mode sector, we expect that physical
quantities should be independent of the specific $\vp$ we choose (\ie,
for the family of solutions defined by eq. \(gencompgp) physical
quantities should be independent of the parameter $\a$).  It may be,
however, that particular choices for $\vp$ lead to particularly simple
formulations.  It is instructive in this regard to examine the naively
kinematical Poincar\'e generators $P^+$ and $P^i$, and check whether
they remain kinematical when the zero mode contributions are included.

The operators $P^\mu$ are defined by
$$
P^\mu = \ha \int dx^-d^2x_\perp\ths T^{+\mu}\; ,
					\eqno(pincaregens)
$$
where we take for the energy-momentum tensor the gauge-invariant form
$$
T^{\mu\nu} =-F^{\mu\l}F^\nu_{~\l}-g J^\mu A^\nu
+i\psibar\g^\mu\del^\nu\psi - g^{\mu\nu}{\cal L}
					\eqno(basicemt)
$$
with ${\cal L}$ given in eq. \(lagrangian).  Eq. \(basicemt) differs
by the addition of a total divergence from what we obtain by a
straightforward application of Noether's theorem. For $P^+$ the
relevant component is
$$
T^{++} = 4(\del_-\nm{A^i})^2 + 4(\del_-\nm{A^i})(\del_i\zm{A}^+)
+(\del_i\zm{A}^+)^2 - gJ^+\zm{A}^+ + 4i\psi_+^\dagger\del_-\psi_+\; .
					\eqno(tplusplus)
$$
The first and last terms in \(tplusplus) just give the usual normal
mode contribution to $P^+$.  The second term vanishes upon integration
in $x^-.$  Finally, the remaining two terms combine, after a transverse
integration by parts, to give a contribution to $P^+$
$$
\ha\int dx^-d^2x_\perp\ths\zm{A}^+(-\del_\perp^2\zm{A}^+
-g\zm{J}^+)\; .
					\eqno(nastyjunk)
$$
This vanishes upon implementing the constraint \(zm2).  Thus $P^+$
remains kinematical, even with the zero modes present.

For the $P^i$ we require
$$
T^{+i}=-(\del_+\zm{A}^+)(\del_i\zm{A}^+)
-(\del_j\zm{A}^+)(\del_i\zm{A}^j)+(\del_j\zm{A}^+)(\del_j\zm{A}^i)
-g\zm{J}^+\zm{A}^i+\dots
					\eqno(tplusi)
$$
where we have omitted the purely normal mode contributions and terms
that vanish upon integration.  The last two terms cancel upon
integration by parts and application of the constraint \(zm2).  The
first two combine to give a contribution to $P^i$
$$
-\ha\int dx^-d^2x_\perp (\del_i\zm{A}^+)
(\del_+\zm{A}^+ + \del_j\zm{A}^j)\; .
					\eqno(zmcontrib)
$$
Clearly, $P^i$ will contain zero mode contributions, and hence will be
``interacting,'' unless
$$
\del_+\zm{A}^+ + \del_j\zm{A}^j = 0\; .
					\eqno(bestgauge)
$$
This corresponds to taking $\a=0$ in \(aizeromode).  Interestingly,
this condition amounts to a zero mode projection of the {\it Lorentz}
gauge condition
$$
\del_\mu \zm{A}^\mu = 0
\;.
					\eqno(zmlorentz)
$$

What does it mean for the $P^i$ to not have the same form as in free
field theory?  In this case it will be impossible to realize the
Heisenberg equation
$$
[\psi_+,P^i]=-i\del_i\psi_+
					\eqno(perpheisenbergs)
$$
with a {\it simple} anticommutation relation between $\psi_+$ and
$\psi_+^\dagger$.  In order to obtain eq. \(perpheisenbergs) through
${\cal O}(g^2)$ we would have to take
$$
\{\psi_+(\ub{x}),\psi_+^\dagger(\ub{x}^\pri)\}
=\L_+[\d^{(3)}(\ub{x}-\ub{x}^\pri) +{\cal O}(g^2\a)+\dots]\; ,
					\eqno(yuckycomm)
$$
with the ${\cal O}(g^2\a)$ term in \(yuckycomm) chosen so that the
part of $[\psi_+,P^i]$ coming from the interacting terms in $P^i$ is
canceled by contribution from the free-field part of $P^i$ and the
${\cal O}(g^2\a)$ piece of the anticommutator.

In fact, we could determine the required anticommutation relation as
follows.  Consider the theory with $\a=0,$ where since the $P^i$ have
their usual free-field forms the standard canonical anticommutator for
$\psi_+$ is the correct one.  Now perform a redefinition of $\psi_+$
that corresponds to the gauge transformation that would take us from
$\a=0$ to $\a\neq0$, specifically
$$
\psi_+=e^{-ig\L}\psi_+^\pri
					\eqno(redefn)
$$
with
$$
\L=\a g{1\over(\del_\perp^2)^2}\del_j \zm{J}^j\; .
					\eqno(redefndefn)
$$
It is straightforward to check that when written in terms of
$\psi_+^\pri,$ the $P^\mu$ have the same forms they would have if we
had started with $\a\neq0$ in eq. \(aizeromode).  In particular, the
$P^i$ acquire a term equal to \(zmcontrib).$^{\ddagger4}$
\footnote{}{$^{\ddagger4}$ We are being somewhat cavalier here about
issues of operator ordering, \etc, which affect the precise form
of the field redefinition \(redefn).}
Thus $\psi_+^\pri$ satisfies the commutation relations necessary to
obtain eq. \(perpheisenbergs) for $\a\neq0,$ and we can simply compute
these by inverting the redefinition \(redefn) and using the known
commutation relations for $\psi_+.$ All of this is not really
necessary, however.  The point is that different values of $\a$ are
physically equivalent; they are just related by redefinitions of
$\psi_+.$ What is special about $\a=0$ (or more generally $\vp=0$ in
eq. \(secsoln)) is that this is the unique choice for which simple
commutation relations among the fields are possible.  Thus it is most
sensible to take $\a=0$ from the very beginning, and for the remainder
of the paper this is what we shall do.

It is perhaps also worth noting that $\a=0$ also results in the
simplest constraint relation for $\zm{A}^-$ (see eq. \(solvingam0) and
the discussion following eq. (3.1)).  Indeed, in this case all of the
constrained zero modes satisfy
$$
\zm{A}^\mu = -g{1\over\del_\perp^2}\zm{J}^\mu\; ,
					\eqno(symmetricsoln)
$$
which has a pleasing symmetry.

Our next task is to solve the constraint relations for the determined
fields and construct the dynamical operators.  As a prelude to the
next section let us briefly remark that because the transverse
currents themselves depend on $\zm{A}^i,$ the structure of
eq. \(aizeromode) is somewhat more complicated than a first glance
reveals.  Nonperturbative solutions to these constraints have so far
proven difficult to obtain.  Nevertheless, an important first step
towards understanding the implications of the zero modes is to examine
them in perturbation theory.  Thus we shall now pursue a perturbative
solution for the $\zm{A}^i$.  This is equivalent to a Fredholm
iterative treatment\r{kp94}.

\subhead{3. Perturbative Formulation}
\taghead{3.}

We now wish to solve the constraint relations \(zm2solved),
\(aizeromode), and \(solvingam0) for the zero modes, and compute
the dynamical operators of the theory.  The components of
$J^\mu$ are given in terms of $\psi_\pm$ by
$$
\eqalignno{
J^\pm & = 2\psi^\dagger_\pm\psi_\pm &(currents a)\cr J^i &=
\psi^\dagger_+ \a^i \psi_- + \psi^\dagger_-\a^i \psi_+\; .&(currents b)
}
$$
{}From these and an inspection of the constraint equation \(psimeom)
for $\psi_-$ we can easily identify which zero modes are simple
and which are difficult to compute.  The field $\zm{A}^+$ is trivially
obtained from eq. \(zm2solved), since it depends only on the dynamical
part of the fermi field $\psi_+.$ The transverse fields $\zm{A}^i$ are
more complicated, since $\zm{J}^i$ depends on both $\zm{A}^i$ and
$\zm{A}^+.$ Thus eq. \(aizeromode) actually determines $\zm{A}^i$
implicitly, and obtaining a general solution for $\zm{A}^i$ is quite
difficult in the quantum theory.  In some sense this is to be
expected, however: the complexity normally associated with the vacuum
state when quantizing on a spacelike surface has to go somewhere.
Finally, $\zm{A}^-$ will be as difficult to determine as the
$\zm{A}^i$ unless $\a=0$; using eq. \(psipeom) to express the
$x^+$-derivative of $J^i$ in terms of the fields on $x^+=0$ will
introduce $\zm{A}^-$ into the right hand side of eq. \(solvingam0).
If $\a=0,$ however, then eq. \(solvingam0) allows a straightforward
computation of $\zm{A}^-$ in terms of $\psi_+$, $\nm{A^i}$, and
$\zm{A}^i$.  In this case the only approximations necessary to
calculate $\zm{A}^-$ are those needed for the computation of $\zm{A}^i.$

We shall now construct a perturbative solution to eq. \(aizeromode)
and study the structure of the theory to lowest nontrivial order.
This requires constructing the Hamiltonian through terms of
${\cal O}(g^2),$ which in turn corresponds to taking the
${\cal O}(g)$ solution for $\zm{A}^i.$ We obtain this simply by setting
$g=0$ in the current $\zm{J}^i$, or, in other words, by using the
zeroth-order expression for $\psi_-$:
$$
\psi_-^{(0)} = {1\over2{i\partial_-}}(-i\a^i \partial_i+ m\b) \psi_+\; .
					\eqno(treepsim)
$$
Inserting the Fourier expansion of the field $\psi_+$ we then find
that the (normal-ordered) proper zero mode of the transverse current
is given to this order by
$$
\eqalignno{
:{\zm{J}^i}: &=
-{2\over\Omega} {\sum_{s,\ub{k},\ub{k}^\pri}}^\pri
\, {\d_{k^+,{k^\pri}^+}\over k^+}
\Bigl[\e^i_{2s}({\e}_{-2s}\cdot k^\pri_\perp)
	+\e^i_{-2s}({\e}_{2s}\cdot k_\perp)\Bigr] &(jperpz)\cr
&\qquad\qquad\qquad
\times \Bigl[b^\dagger_{s\ukp} b_{s\uk}
e^{-i(k_\perp^\pri-k_\perp)\cdot x_\perp}
-d^\dagger_{-s\uk} d_{-s\ukp}
e^{+i(k_\perp^\pri-k_\perp)\cdot x_\perp}
\Bigr]\; ,
}
$$
where the prime on the sum indicates that terms with
$k_\perp=k_\perp^\pri$ are to be excluded (\ie\ the global zero mode
is removed as per eq. \(pzmdef)).  We then obtain $\zm{A}^i$ at ${\cal
O}(g)$ by inserting \(jperpz) into eq. \(aizeromode).  Of course,
$\zm{A}^+$ is obtained simply by substituting the expansion of
$\psi_+$ into eq. \(zm2solved).  Neither of these expressions for the
zero modes themselves is particularly illuminating, however, and we do
not display them.

With the $\zm{A}^i$ and $\zm{A}^+$ in hand we can now construct the
Hamiltonian through ${\cal O}(g^2)$.  For this we need
$$
\eqalignno{
T^{+-} & = (\del_-A^-)^2+(\del_+\szm{A^+})^2+ {1\over2} F^{ij}F^{ij}
+g(J^-\szm{A^+} - 2J^iA^i)
-4\psi_-^\dagger(i\del_-\psi_-)  \cr
&\qquad\qquad  + 2\psi_-^\dagger(-i\a^i\del_i + m\b)\psi_+
+{\rm h.c.}\; ,&(tplusminus)
}
$$
where we have discarded terms that will not contribute when
integrated.  Now, the contribution from the normal mode part of the
theory may be found in various discussions of QED in DLCQ, for example
in refs.\r{tbp91,tangphd}.  In Appendix B we discuss the contributions
from the global zero modes, and show they are unnecessary to this
order at least.  Thus we display here only the parts of $P^-$ arising
from the proper zero modes:
$$
\eqalignno{
P^-_Z & = \ha \int dx^-d^2x_\perp
\Bigl[
{1\over2} (\partial_i \zm{A}^j)^2 - {1\over2} (\partial_i \zm{A}^i)^2
+ (\del_+\zm{A}^+)^2
- 2g \zm{J}^i \zm{A}^i
+ g \zm{J}^- \zm{A}^+ &(pminusZ)\cr
& \quad
-4\psi_-^{(0) \dagger} (i\del_-\psi_-^{(2)})
-4\psi_-^{(2) \dagger} (i\del_-\psi_-^{(0)})
+ 2\psi_-^{(2)\dagger}(-i\a^i\del_i + m\b)\psi_+
+{\rm h.c.} \Bigr]\; .
}
$$
Here $\psi_-^{(2)}$ is the second-order correction to the dependent
fermi field, which comes entirely from the zero modes:
$$
\psi_-^{(2)} = -g\a^i\szm{A^i} {1\over2i\del_-} \psi_+
+g\szm{A^+}{1\over(2i\del_-)^2}(-i\a^i\del_i+m\b)\psi_+\; .
					\eqno(whatpsi2is)
$$
It turns out that all the terms in eq. \(pminusZ) involving
$\psi_-^{(2)}$ cancel among themselves.  Furthermore, after
implementing the constraints the terms
$$
\ha \int dx^-d^2x_\perp
\Biggl[
{1\over2} (\partial_i \zm{A}^j)^2 - {1\over2} (\partial_i \zm{A}^i)^2
+ (\del_+\zm{A}^+)^2
- 2g \zm{J}^i \zm{A}^i \Biggr]
					\eqno(someterms)
$$
in eq. \(pminusZ) may be combined to give
$$
{g^2\over2} \int dx^- d^2x_\bot
\Biggl[ \zm{J}^i {1\over{\del_\bot^2}} \zm{J}^i \Biggr]
\; .
					\eqno(rewritten)
$$
Thus the complete contribution to $P^-$ at this order from the proper
zero modes reduces to
$$
P^-_Z = {g^2\over2} \int dx^- d^2x_\bot
\Biggl[ \zm{J}^i {1\over{\del_\bot^2}} \zm{J}^i
-\zm{J}^-{1\over\del_\perp^2}\zm{J}^+ \Biggr]\; ,
					\eqno(zmcontri)
$$
where we have used eq. \(zm2) to express $\zm{A}^+$ in terms of
$\zm{J}^+.$

It is now a straightforward (if tedious) exercise to insert the
Fourier expansion of $\psi_+$ into eq. \(zmcontri) and express $P^-_Z$
in the Fock representation.  We obtain new four-fermion operators, as
well as fermion bilinears which arise when the four-point terms are
brought into normal order (as usual, $c$-number contributions that
result are discarded so that $\langle0|P^-|0\rangle=0$).  These latter
terms have been called ``self-induced inertias'' in the literature,
since they have the Fock space structure of a mass term.  One final
comment is warranted before we present the explicit terms.  The
operators $\zm{J}^-$ and $\zm{J}^+$ in eq. \(zmcontri) do not commute.
Therefore the last term in eq. \(zmcontri) is non-Hermitian as it
stands.  This operator ordering ordering ambiguity is treated by
symmetrization:
$$
\zm{J}^- {1\over\del_\perp^2} \zm{J}^+ \rightarrow
{1\over2} \Bigl(
\zm{J}^- {1\over\del_\perp^2} \zm{J}^+
+ \zm{J}^+ {1\over\del_\perp^2} \zm{J}^- \Bigr)\;.
					\eqno(symmetrization)
$$

The results are conveniently grouped into four sets of four-fermion
operators and the self-inertias:
$$
P^-_Z=P^-_T + P^-_L + P^-_{m} + P^-_{m^2} + P^-_{sii}\; ,
					\eqno(grouping)
$$
where $P^-_T$ and $P^-_L$ are the $m$-independent contributions from
the first and second terms in eq. \(zmcontri), respectively, $P^-_{m}$
and $P^-_{m^2}$ are all contributions proportional to $m$ and $m^2,$
respectively, and $P^-_{sii}$ is the full self-inertia
contribution.  We find
$$
\eqalignno{
P^-_T & = {g^2\over\Omega}
{\sum_{s,\uk,\ukp}}^\pri{\sum_{t,\up,\upp}}^\pri
{ \d_{k^+,{k^\pri}^+} \d_{p^+,{p^\pri}^+}\over
k^+p^+(k_\perp-k^\pri_\perp)^2
(p_\perp-p^\pri_\perp)^2}&(pminusperp) \cr
&\quad \times
2(k_\perp-k^\pri_\perp)\cdot(p_\perp-p^\pri_\perp)\e^i_{2s} \e^j_{-2s}
\Bigl[ (k^i{p^\pri}^j + {k^\pri}^j p^i) \d_{s,t}
+(k^i p^j + {k^\pri}^j {p^\pri}^i) \d_{s,-t}
\Bigr]\cr
&\qquad \times\Bigl[
b^\dagger_{s\ukp} b^\dagger_{t\upp} b_{s\uk} b_{t\up}
\ths\d^{(2)}_{k^\pri_\perp-k_\perp,p_\perp-p^\pri_\perp}
+b^\dagger_{s\ukp}d^\dagger_{-t\up}b_{s\uk}d_{-t\upp}
\ths\d^{(2)}_{k^\pri_\perp-k_\perp,p^\pri_\perp-p_\perp} \cr
& \qquad\quad
+d^\dagger_{-s\uk}b^\dagger_{t\upp}d_{-s\ukp}b_{t\up}
\ths\d^{(2)}_{k^\pri_\perp-k_\perp,p^\pri_\perp-p_\perp }
+d^\dagger_{-s\uk}d^\dagger_{-t\up}d_{-s\ukp}d_{-t\upp}
\ths\d^{(2)}_{k^\pri_\perp-k_\perp,p_\perp-p^\pri_\perp}
\Bigr]
}
$$

$$
\eqalignno{
P^-_L & = {4g^2\over\Omega}
\sum_{s,\uk,\ukp}{\sum_{t,\up,\upp}}^\pri
{\d_{k^+,{k^\pri}^+} \d_{p^+,{p^\pri}^+}\over (k^+)^2}
\ths{\e^i_{2s} \e^j_{-2s} k^i {k^\pri}^j \over
(p_\perp-p^\pri_\perp)^2} &(pminuslong)\cr
&\qquad \times\Bigl[
b^\dagger_{s\ukp}b^\dagger_{t\upp}b_{s\uk}b_{t\up}
\ths\d^{(2)}_{k^\pri_\perp-k_\perp,p_\perp-p^\pri_\perp}
-b^\dagger_{s\ukp}d^\dagger_{-t\up}b_{s\uk}d_{-t\upp}
\ths\d^{(2)}_{k^\pri_\perp-k_\perp,p^\pri_\perp-p_\perp} \cr
& \qquad\quad
-d^\dagger_{-s\uk}b^\dagger_{t\upp}d_{-s\ukp}b_{t\up}
\ths\d^{(2)}_{k^\pri_\perp-k_\perp,p^\pri_\perp-p_\perp}
+d^\dagger_{-s\uk}d^\dagger_{-t\up}d_{-s\ukp}d_{-t\upp}
\ths\d^{(2)}_{k^\pri_\perp-k_\perp,p_\perp-p^\pri_\perp}
\Bigr]
}
$$

$$
\eqalignno{
P^-_{m} & = {2\sqrt{2}g^2m\over\Omega}\sum_{s,\uk,\ukp}
{\sum_{t,\up,\upp}}^\pri
{\d_{k^+,{k^\pri}^+} \d_{p^+,{p^\pri}^+}\over (k^+)^2}
\ths{\e^i_{2s}(k^i-{k^\pri}^i) \over
(p_\perp-p^\pri_\perp)^2}&(pminusm)\cr
&\qquad \times\Bigl[
+b^\dagger_{-s\ukp}b^\dagger_{t\upp}b_{s\uk}b_{t\up}
\ths\d^{(2)}_{k^\pri_\perp-k_\perp,p_\perp-p^\pri_\perp}
-b^\dagger_{-s\ukp}d^\dagger_{-t\up}b_{s\uk}d_{-t\upp}
\ths\d^{(2)}_{k^\pri_\perp-k_\perp,p^\pri_\perp-p_\perp} \cr
& \qquad\quad
+d^\dagger_{-s\uk}b^\dagger_{t\upp}d_{s\ukp}b_{t\up}
\ths\d^{(2)}_{k^\pri_\perp-k_\perp,p^\pri_\perp-p_\perp}
-d^\dagger_{-s\uk}d^\dagger_{-t\up}d_{s\ukp}d_{-t\upp}
\ths\d^{(2)}_{k^\pri_\perp-k_\perp,p_\perp-p^\pri_\perp}
\Bigr]
}
$$

$$
\eqalignno{
P^-_{m^2} & = {2g^2m^2\over\Omega}\sum_{s,\uk,\ukp}
{\sum_{t,\up,\upp}}^\pri
{\d_{k^+,{k^\pri}^+} \d_{p^+,{p^\pri}^+}\over (k^+)^2}
{1\over (p_\perp-p^\pri_\perp)^2} &(pminusmm)\cr
&\qquad \times\Bigl[
-b^\dagger_{s\ukp}b^\dagger_{t\upp}b_{s\uk}b_{t\up}
\ths\d^{(2)}_{k^\pri_\perp-k_\perp,p_\perp-p^\pri_\perp}
+b^\dagger_{s\ukp}d^\dagger_{-t\up}b_{s\uk}d_{-t\upp}
\ths\d^{(2)}_{k^\pri_\perp-k_\perp,p^\pri_\perp-p_\perp} \cr
&\qquad\quad
+d^\dagger_{-s\uk}b^\dagger_{t\upp}d_{-s\ukp}b_{t\up}
\ths\d^{(2)}_{k^\pri_\perp-k_\perp,p^\pri_\perp-p_\perp}
-d^\dagger_{s\uk}d^\dagger_{t\up}d_{s\ukp}d_{t\upp}
\ths\d^{(2)}_{k^\pri_\perp-k_\perp,p_\perp-p^\pri_\perp}
\Bigr]
}
$$

$$
\eqalignno{
P^-_{sii} & = {g^2\over\Omega}{\sum_{s,\uk,\up}}^\pri
	{\d_{k^+,p^+}\over(k^+)^2(k_\perp-p_\perp)^2}
\Bigl[2m^2 - (k_\perp^2+p_\perp^2)
-2\e^i_{2s}\e^j_{-2s} (k^ip^j+k^jp^i) \Bigr] &(pminussii)\cr
&\qquad\qquad\qquad\qquad
\times \Bigl[
b^\dagger_{s\up}b_{s\up} + d^\dagger_{s\up}d_{s\up}
\Bigr]\; .
}
$$
For completeness we also display $P^+$ and $P^i$, which for $\a=0$
have the usual kinematical forms:
$$
P^+ = \sum_{s,\up} p^+\Bigl[b^\dagger_{s\up}b_{s\up} +
d^\dagger_{s\up}d_{s\up}\Bigr]
+ \sum_{\l,\uk} k^+ \ths a^\dagger_{\l\uk}a_{\l\uk}
					\eqno(pplus)
$$

$$
P^i = \sum_{s,\up} p^i\Bigl[b^\dagger_{s\up}b_{s\up} +
d^\dagger_{s\up}d_{s\up}\Bigr]
+ \sum_{\l,\uk} k^i \ths a^\dagger_{\l\uk}a_{\l\uk}\; .
					\eqno(pperp)
$$

It is straightforward to check that through ${\cal O}(g^2)$ all
Heisenberg equations reduce to the appropriate field equations, and
that the Poincar\'e algebra is correctly obtained.  We therefore have
a valid representation of the dynamics defined by the equations of
motion \(psipeom)--\(aperpeom).

\subhead{4. The Fermion Self-Energy}
\taghead{4.}

We shall now examine the effects of the zero mode contributions on the
one-loop fermion self-energy in this theory.  In ref.\r{mr92} it was
found that the zero mode contributions to $P^-$ included a counterterm
that removed a certain noncovariant, quadratic divergence in the
fermion self-energy (eigenvalue of $P^-$) in a Yukawa theory.  We wish
to see whether the same thing happens in QED.

The fermion self-energy is not the only quantity to which $P^-_Z$
contributes at ${\cal O}(g^2)$, of course.  The various four-fermion
operators in $P^-_Z$ will certainly contribute to tree-level
scattering amplitudes.  There will also be divergent contributions to
the $e^+e^-\g$ vertex, and hence to the charge renormalization, at
lowest order.  A complete discussion of the effects of the new terms
in $P^-$ on the one-loop renormalization of this theory will be
presented elsewhere\r{krinprep}.

Let us first discuss the contributions to the fermion self-energy
coming from the normal mode sector of the theory.  These can
essentially be taken from the work of Tang\r{tbp91,tangphd}, with one
caveat to be mentioned below.  There are two contributions, one coming
from one-fermion-one-photon intermediate states ($\d P^-_1$), and one
coming from the self-inertia terms in the Hamiltonian ($\d P^-_2$).
We find
$$
\d P^-_1=-{\a L\over\pi L_\perp^2}\sum_{\senk{q}}
\sum_{q=2,4,\dots}^{n-1} {{1\over n(n-q)}
\Bigl[n^2(\senk{q}-{q\over n}\senk{n})^2+q^2\b_f\Bigr]
+{2n^2\over q^2}(\senk{q}-{q\over n}\senk{n})^2
\over n^2(\senk{q}-{q\over n}\senk{n})^2 + q^2\b_f}
					\eqno(deltae1)
$$

$$
\d P^-_2={\a L\over\pi L_\perp^2}\sum_{\senk{q}}
\Biggl\{\sum_{m=1,3,\dots}^\infty{1\over (n-m)^2}-
{1\over (n+m)^2}
+\ha \sum_{q=2,4,\dots}^\infty
{1\over q(n-q)} + {1\over q(n+q)}\Biggr\}
					\eqno(deltae2)
$$
where $({n\pi\over L},{\senk{n}\pi\over L_\perp})$ is the momentum of
the fermion, $\a\equiv {g^2\over 4\pi},$ and
$\b_f\equiv\Bigl({mL_\perp\over\pi}\Bigr)^2.$ The caveat is that in
obtaining eq. \(deltae2) we have symmetrized the self-inertias given
in refs.\r{tbp91,tangphd} under $b\lra d$.  This is the effect of
using the explicitly $C$-odd form of the current
$$
J^\mu=\ha[\psibar,\g^\mu\psi]
					\eqno(realcurrent)
$$
(see for example\r{bandd}), or equivalently of properly symmetrizing
products of noncommuting operators in the construction of
$P^-$\r{mrunpub}.

Now, each of these contributions is separately quadratically divergent
in a transverse momentum cutoff.  In a continuum formulation, with a
suitable (Lorentz covariant) regulator, these quadratic divergences
cancel and we recover the expected logarithmic singularity.  Here,
however, the coefficient of ${\a L\over2\pi L_\perp^2}\sum_{\senk{q}}$
for $|\senk{q}\ths|\ra\infty$ is
$$
\eqalignno{
\Delta_n & \equiv
-2\sum_{q=2,4,\dots}^{n-1}\Biggl[{1\over n(n-q)}
+ {2\over q^2}\Biggr]
+2\sum_{m=1,3,\dots}^\infty \Biggl[{1\over(n-m)^2}
- {1\over(n+m)^2}\Biggr]\cr
&\qquad\qquad\qquad\qquad\qquad
+\sum_{q=2,4,\dots}^\infty\Biggl[{1\over q(n-q)}
+{1\over q(n+q)}\Biggr]\; .&(2)
}
$$
The sums in eq. \(2) may be evaluated explicitly to give
$$
\Delta_n = {1\over n^2} - {2\ln 2\over n}\; .
					\eqno(deltareduced)
$$
Thus without the zero modes the fermion self-energy contains a
quadratically divergent piece proportional to ${1\over(p^+)^2}.$ This
would of course need to be removed by a counterterm, but one which
cannot correspond simply to the redefinition of a parameter in the
Lagrangian.  In contrast, the part proportional to ${1\over p^+}$ can
be interpreted as a correction to the fermion mass.

Finally, let us consider the contribution from the new self-inertia
terms \(pminussii), which is the sole effect of the zero modes at
this order.  These give
$$
\d P^-_0={g^2\over\Omega}{1\over(p^+)^2}\sum_{k_\perp}
{1\over(k_\perp-p_\perp)^2}\Bigl[
2m^2-(k_\perp^2+p_\perp^2)
-2\e^i_{2s}\e^j_{-2s} (k^ip^j + k^jp^i)
\Bigr]\; ,
					\eqno(basicnew)
$$
which is quadratically divergent for $|k_\perp|\ra\infty$:
$$
\delta P^-_0\approx -
{\a L\over2\pi L_\perp^2}{1\over n^2}\sum_{k_\perp}\; .
					\eqno(zmcont)
$$
The corresponding correction to $\Delta_n$ is therefore $-{1\over
n^2}$, so that the noncovariant part of the quadratic divergence is in
fact cancelled when the zero modes are included.  The quadratic
divergence proportional to ${1\over p^+}$ survives, unlike in the
continuum, but this may be removed by a redefinition of the fermion
mass.  Its occurrence can presumably be traced to the violation of
parity that is a generic feature of DLCQ\r{mr92}.

\subhead{5. Discussion}
\taghead{5.}

We have shown how to perform a general gauge-fixing of Abelian gauge
theory in DLCQ, and cleanly separate the dynamical from the
constrained zero-longitudinal momentum fields.  The various zero mode
fields {\it must} be retained in the theory if the equations of motion
are to be realized as the Heisenberg equations.  We have further seen
that taking the constrained fields properly into account renders the
ultraviolet behavior of the theory more benign, in that it results in
the automatic generation of a counterterm for a noncovariant
divergence in the fermion self-energy in lowest-order perturbation
theory.  Additional effects of the zero mode contributions to $P^-,$
for example on the charge renormalization, are currently under
study\r{krinprep}.

The solutions to the constraint relations for the $\zm{A}^i$ are all
physically equivalent, being related by different choices of gauge in
the zero mode sector of the theory.  There is a gauge which is
particularly simple, however, in that the fields may be taken to
satisfy the usual canonical anticommutation relations.  This is most
easily exposed by examining the kinematical Poincar\'e generators, and
finding the solution for which these retain their free-field forms.
The unique solution that achieves this is $\vp=0$ in eq. \(secsoln).
For solutions other than this one, complicated commutation relations
between the fields will be necessary to correctly translate them in
the initial-value surface.

It would be interesting to study the structure of the operators
induced by the zero modes from the point of view of the light-cone
power-counting analysis of Wilson\r{wilsonunpub,wwhzpg94}.  As noted
in the Introduction, to the extent that DLCQ coincides with reality,
effects which we would normally associate with the vacuum must be
incorporated into the formalism through the new, noncanonical
interactions arising from the zero modes.  Particularly interesting is
the appearance of operators that are nonlocal in the transverse
directions (eq. \(zmcontri)).  These are interesting because the
strong infrared effects they presumably mediate could give rise to
transverse confinement in the effective Hamiltonian for QCD.  There is
longitudinal confinement already at the level of the canonical
Hamiltonian; that is, the effective potential between charges
separated only in $x^-$ grows linearly with the separation.  This
comes about essentially from the nonlocality in $x^-$ (\ie\ the
small-$k^+$ divergences) of the light-cone formalism.

It is clearly of interest to develop nonperturbative methods for
solving the constraints, since we are ultimately interested in
nonperturbative diagonalization of $P^-.$ Several approaches to this
problem have recently appeared in the literature\r{hksw92,bpv93,pv94},
in the context of scalar field theories in 1+1 dimensions.  For QED
with a realistic value of the electric charge, however, it might be
that a perturbative treatment of the constraints could suffice; that
is, that we could use a perturbative solution of the constraint to
construct the Hamiltonian, which would then be diagonalized
nonperturbatively.  An approach similar in spirit has been proposed in
ref.\r{wwhzpg94}, where the idea is to use a perturbative realization
of the renormalization group to construct an effective Hamiltonian for
QCD, which is then solved nonperturbatively.  There is some evidence
that this kind of approach might be useful.  Wivoda and Hiller have
recently used DLCQ to study a theory of neutral and interacting
charged scalar fields in 3+1 dimensions\r{wh93}.  They discovered that
including four-fermion operators precisely analogous to the
perturbative ones appearing in $P^-_Z$ significantly improved the
numerical behavior of the simulation.

The extension of the present work to the case of QCD is complicated by
the fact that the constraint relations for the gluonic zero modes are
nonlinear, as in the $\p^4$ theory.  A perturbative solution of the
constraints is of course still possible, but in this case, since the
effective coupling at the relevant (hadronic) scale is large, it is
clearly desirable to go beyond perturbation theory.  In addition,
because of the central role played by gauge fixing in the present
work, we may expect complications due to the Gribov
ambiguity\r{gribov78}, which prevents the selection of unique
representatives on gauge orbits in nonperturbative treatments of
Yang-Mills theory.  As a preliminary step in this direction, work is
in progress on the pure glue theory in 2+1 dimensions\r{kpp94}. There
it is expected that some of the nonperturbative techniques used
recently in 1+1 dimensions\r{bpv93,pv94} can be applied.

\subhead{ACKNOWLEDGEMENTS}

\noindent
It is a pleasure to thank G. McCartor, H.-C. Pauli, R.J. Perry, S.S.
Pinsky, S. Elser, and F. W\"olz for helpful discussions.
ACK is supported by the DFG under contract DFG-Gz:Pa 450/1-1 and by
NATO under Collaborative Research Grant No. CRG930072, and thanks the
Department of Physics of the Ohio State University for its hospitality
during a visit where some of this work was conducted.
DGR is supported by the National Science Foundation under Grant
Nos. PHY-9203145, PHY-9258270, PHY-9207889, and PHY-9102922.

\subhead{Appendix A: Notation}
\taghead{A.}

\itsubhead{A.1 Light-Cone Coordinates}

We define $x^\pm\equiv x^0\pm x^3$ and take $x^+$ to be the evolution
parameter.  We use latin indices $(i,j,\dots)$ to index transverse
components $x_\perp=(x^1,x^2)$.  A contraction of four-vectors
decomposes as $A\cdot B =\ha(A^+B^- + A^-B^+) - A^i B^i,$ from which
we infer the metric $g_{+-}=g_{-+}=\ha,$ $g^{11}=g^{22}=-1,$ with all
other components vanishing.  Derivatives are defined by
$\del_\pm\equiv\del/\del x^\pm$, $\del_i\equiv\del/\del x^i.$

We shall also make use of an underscore notation: for position-space
variables we write $\ub{x} \equiv (x^-,x_\perp),$ while for
momentum-space variables $\ub{k} \equiv (k^+,k_\perp).$  Then $\ub{k}
\cdot\ub{x} \equiv\ha k^+x^- - k_\perp\cdot x_\perp.$

We further employ Dirac's notation
$$
\a^i\equiv\g^0\g^i\qquad\qquad\b\equiv\g^0\; ,
					\eqno(diracnot)
$$
and define $\g^\pm\equiv\g^0\pm\g^3.$  The Hermitian operators
$$
\L_\pm\equiv\ha\g^0\g^\pm
					\eqno(lambdapm)
$$
serve to project out the constrained and dynamical components of the
fermi field:
$$
\psi_\pm\equiv\L_\pm\psi\; .
					\eqno(psiprojection)
$$
In the Dirac representation of the $\g$-matrices
$$
\L_+=\ha\left(\matrix{1&0&1&0\cr 0&1&0&-1\cr
		      1&0&1&0\cr 0&-1&0&1\cr}\right)\; ,
					\eqno(whatlplusis)
$$
which has two eigenvectors, both with eigenvalue $1$:
$$
\chi_{+\ha}={1\over\sqrt2}\left(\matrix{ 1\cr0\cr1\cr0\cr}\right) \qquad\qquad
\chi_{-\ha}={1\over\sqrt2}\left(\matrix{ 0\cr1\cr0\cr-1\cr}\right)\; .
					\eqno(whatchiis)
$$
These serve as a convenient spinor basis for the expansion of $\psi_+$
on the light cone.

\itsubhead{A.2 Field Expansions and Commutation Relations}

The mode expansions of the fields on $x^+=0$ take the form
$$
\psi_+(\ub{x}) = {1\over\sqrt\Omega}\sum_{s,\uk}
\chi_s\Bigl(b_{s\uk}e^{-i\uk\cdot\ub{x}}
	+ d^\dagger_{-s\uk}e^{i\uk\cdot\ub{x}}\Bigr)
					\eqno(psiexpansion)
$$
$$
\nm{A^i}(\ub{x}) = {1\over\sqrt\Omega}\sum_{\l,\uq}
\e^i_\l\Bigl(a_{\l\uq}e^{-i\uq\cdot\ub{x}}
	+ a_{\l\uq}^\dagger e^{i\uq\cdot\ub{x}}\Bigr)\; ,
					\eqno(aiexpansion)
$$
where $\Omega\equiv 8LL^2_\perp$ is the spatial volume, the spinors
$\chi_s$ are given in eq. \(whatchiis), and the polarization vectors
$\e^i_\l$ are defined by
$$
\e^i_{+1} = {-1\over\sqrt{2}}(1,i)\qquad\qquad
\e^i_{-1} = {1\over\sqrt{2}}(1,-i)\; .
					\eqno(epsilons)
$$
These satisfy
$$
{\e^i_\l}^*\e^i_{\l^\prime}=\d_{\l\l^\prime} \qquad\quad
\sum_\l \e^i_\l{\e^j_\l}^*=\d^{ij} \quad\qquad
{\e^i_\l}^*=-\e^i_{-\l}\; .
					\eqno(epsrelations)
$$
A useful relation satisfied by the $\chi_s$ and $\e^i_\l$ is
$$
\chi^\dagger_{s^\pri}\a^i\a^j\chi_s =
-2\e^i_{-2s}\e^j_{2s} \d_{s,s^\pri}\; ;
					\eqno(identity)
$$
others may be found in ref.\r{tangphd}.  Recall that the gauge field
is taken to be periodic in all coordinates, while the fermi field is
periodic in $x_\perp$ and antiperiodic in $x^-.$ Thus in eq.
\(psiexpansion) the sum runs over the allowed momenta
$$
k^+={n\pi\over L},\quad n=1,3,5,\dots \quad;\quad
k^i={n^i\pi\over L_\perp},\quad n^i=0,\pm1,\pm2,\dots
					\eqno(psimomenta)
$$
while in \(aiexpansion) we have
$$
q^+={m\pi\over L},\quad m=2,4,6,\dots \quad;\quad
q^i={m^i\pi\over L_\perp},\quad m^i=0,\pm1,\pm2,\dots
					\eqno(phimomenta)
$$

The canonical commutation relations to be imposed are
$$
\{\psi_{+\a}(\ub{x}),\psi^\dagger_{+\b}(\ub{x}^\prime)\}
=(\L_+)_{\a\b}\kd3(\ub{x}-\ub{x}^\prime)
					\eqno(psiccr)
$$
$$
[\nm{A^i}(\ub{x}),\del^+\nm{A^j}(\ub{x}^\prime)]
=i\d^{ij}\Bigl[
\kd3(\ub{x}-\ub{x}^\prime)-{1\over2L}\d^{(2)}(x_\perp-x^\pri_\perp)
\Bigr]\; .
					\eqno(phitccr)
$$
These are realized by the Fock space relations
$$
\{b_{s\ub{k}},b^\dagger_{s^\pri\ub{k}^\pri}\}=
\{d_{s\ub{k}},d^\dagger_{s^\pri\ub{k}^\pri}\}=
\d_{ss^\prime}\kd3_{\ub{k},\ub{k}^\pri}
					\eqno(ccrs1)
$$
$$
[a_{\l\ub{q}},a^\dagger_{\l^\pri\ub{q}^\pri}]
=\d_{\l\l^\pri} \kd3_{\ub{q},\ub{q}^\pri}
					\eqno(ccrs2)
$$
$$
\{b,b\}=\{d,d\}=\{b,d^\dagger\}=[a,b]=[a,d]=[a,b^\dagger]=[a,d^\dagger]
=0\; .
					\eqno(ccrs3)
$$

\subhead{Appendix B: The Global Zero Mode Sector}

As discussed in ref.\r{kp94}, the Gauss law in the global zero mode
sector is just the vanishing of the total charge
$$
\totzm{J^+} = 0\; .
					\eqno(qtotiszip)
$$
This is a first-class constraint in the Dirac sense\r{dirac64}.  Thus it
can only be realized as a condition on physical states in the quantum
Hilbert space:
$$
:\totzm{J^+}: |\rm{phys} \rangle = 0\; ,
					\eqno(actuallythis)
$$
where normal ordering eliminates the zero point infinity.  In terms of
the Fock operators
$$
:\totzm{J^+}: = {g \over \Omega} \sum_{s,\uk}
(b_{s \uk}^\dagger b_{s \uk} - d_{s \uk}^\dagger d_{s \uk})\; .
					\eqno(qfockspace)
$$
Consequently, to lowest nontrivial order in the coupling constant $g$,
charge singlet states are eigenstates of the global zero modes of
the remaining source components
$$
\eqalignno{
:\totzm{J^-}: & = {g \over \Omega} \sum_{s,\uk}
{{k^2_\bot + m^2} \over {2k^+}}
(b_{s \uk}^\dagger b_{s \uk} - d_{s \uk}^\dagger d_{s \uk})&(otherjs a)
\cr
:\totzm{J^i}: & = {g \over \Omega} \sum_{s,\uk}
{{k^i} \over {2k^+}}
(b_{s \uk}^\dagger b_{s \uk} - d_{s \uk}^\dagger d_{s \uk})\; .&(otherjs b)
}
$$
This is important for the global sector corresponding to $A^+$.  For
this we obtain the following contribution to the Hamiltonian
$$
P^-_{\rm{glob}} = \Omega \Bigl[ {1\over2}
\totzm{\pi^-}^2 + g \totzm{A^+} \totzm{J^-} \Bigr]\; .
					\eqno(pminusbits)
$$
We observe that $\totzm{A^+}$ represents a genuine degree of freedom
coupling to the electron-photon sector. Thus in lowest-order
perturbation theory the Hilbert space can be constructed in terms of
the product states
$$
\Psi \otimes |\rm{phys}\rangle\; ,
					\eqno(productspace)
$$
where $\Psi$ are stationary wavefunctions satisfying the
Schr\"odinger equation
$$
-{1\over2}{{d^2}\over {dq^2}} \Psi + {g \over \Omega} q v \Psi =
{\cal E} \Psi
					\eqno(schroedinger)
$$
with $q = \Omega \totzm{A^+}$, $v = : \totzm{J^-} :$ and ${\cal E}$
the energy {\it density} $E/\Omega$ .  We cannot solve this exactly,
but perturbation theory in the coupling suffices.  The free particle
$(g=0)$ wavefunctions are
$$
\Psi^{(0)}(q) = {\cal A} e^{i \sqrt{2{\cal E}} q}\; .
					\eqno(freewfns)
$$
There is in fact a boundary condition on $\Psi$. First we observe that
there is a (rather trivial) Gribov ambiguity in the gauge we employ.
Writing the Abelian gauge transformations in the form
$$
A^\mu \rightarrow  U A^\mu U^{-1} + {1 \over {ig}} U \partial^\mu U^{-1}
					\eqno(fancygt)
$$
we see that $U = \exp (i {{2 \pi n} \over L} x^-)$ maintains the gauge
condition $\partial_- A^+ = 0,$ as well as the other conditions, but
shifts the value of the global zero mode by $2\pi n/gL$,
$$
\totzm{A^+} \rightarrow \totzm{A^+} - {{2\pi n} \over {gL}}\; .
					\eqno(shifttzm)
$$
Note that this $U$ is periodic and, due to its $x^-$ dependence, is
{\it not} part of the residual freedom that was left after introducing
the condition $\partial_-A^+ =0$.  Thus the quantum mechanical
particle $q$ transforms as $q \rightarrow q - {{2\pi n} \over {gL}}
\Omega$.  We proceed in analogy with Manton's treatment\r{gaugerefs}
of the Schwinger model on a circle: the wavefunction $\Psi^{(0)}$ is
assigned the boundary condition
$$
\Psi^{(0)}(q) = \Psi^{(0)} (q - {{2\pi n} \over {gL}} ) \; .
					\eqno(boundaryc)
$$
Now we choose the first ``horizon'' corresponding to $n=1$. Within
this we obtain the discrete spectrum of ``$m$-states,''  m an
integer
$$
\Psi^{(0)}_m(q) = {\cal A} e^{i \sqrt{2{\cal E}_m} q},
\quad {\rm{with}} \quad
{\cal E}_m = {{m^2 g^2} \over {8 (2L_\perp)^2} }\; .
					\eqno(discrete)
$$
Observe that, as expected on dimensional grounds, the longitudinal
interval length $L$ has cancelled in the discrete values of the energy
density. In 1+1 dimensions there is no longer any length parameter
left, so even in the naive continuum limit one obtains a finite energy
density. In this case of 3 space dimensions, the naive $L_\perp
\rightarrow \infty$ limit collapses all the $m$-states down to zero
energy. On the other hand, we now show that for lowest-order
perturbation theory, even in the finite volume transitions from any
such state to another excitation are suppressed.

The argument rests on the simplicity of the interaction.  Since the
electron-photon states $|{\rm{phys}}\rangle$ are eigenstates of this
Hamiltonian we can take matrix elements of the Hamiltonian and work
with a reduced Schr\"odinger equation.  Evaluating the first-order
correction to
$$
\langle \Psi^{(0)}_n | \Psi^{(0)}_m \rangle
\equiv \int_{-\infty}^{+\infty} dq \Psi^{(0) \ast}_n(q) \Psi^{(0)}_m (q)
					\eqno(matrixel)
$$
we obtain
$$
g v |{\cal A}|^2 \int_{-\infty}^{+\infty} dq
{q \over {{\cal E}_n - {\cal E}_m}}
e^{i (\sqrt{2{\cal E}_n} - \sqrt{2{\cal E}_m}) q}
\qquad\qquad (n \neq m)\; .
					\eqno(evaluated)
$$
This can be written as the derivative with respect to $\sqrt{2{\cal E}_n} -
\sqrt{2{\cal E}_m}$ of a delta function whose support is {\it empty}, since
$n \neq m$. Thus the correction vanishes, as does the unperturbed
amplitude.  Thus with the system initialized in a given $m$-state, the
interactions will not allow a transition to another state. The effect
is a pure background that can be ignored as far as the electron-photon
theory is concerned.  The reader is referred to ref.\r{kpp94b} for an
examination of the role of such states in pure glue theory in 1+1
dimensions.

Let us next discuss $\totzm{A^i}$.  The problem here is that
projection of the Maxwell equations into the global sector does not
yield information about $\totzm{A^i}$, neither in the form of a
constraint relation nor a dynamical equation of motion.  But the Dirac
theory sees it reintroduced into the equation for $\zm{A}^i$ via
the constraint for $\psi_-$.  Physically, these fields represent
quanta that propagate along the initial-value surface $x^+=0.$ It may
be, therefore, that they should be initialized along a surface
orthogonal to $x^+=0,$ in a way familiar from, \eg\ the treatment of
massless fields in two spacetime dimensions.  This approach is
currently under study.  An alternative treatment has been proposed in
ref.\r{kp94}, in which a mass term is introduced for $\totzm{A^i}$
only.  In this case the global projection of the equation of motion
analogous to eq. \(aperpeom) gives
$$
\mu^2 \totzm{A^i} = g \totzm{J^i}\; ,
					\eqno(newconstraint)
$$
so that $\totzm{A^i}$ becomes constrained.  In this approach it is
simple to check that the contributions to $P^-$ coming from
$\totzm{A^i}$ begin at ${\cal O}(g^3),$ and so would be irrelevant for
the present work.  Lacking a definite prescription for handling these
modes, we shall here simply discard them.  It should be noted that
(unlike the other zero modes) omitting $\totzm{A^i}$ from the theory
does not introduce inconsistencies into the equations of motion.

Finally, the global zero mode $\totzm{A^-}$ is set to zero using the
only gauge freedom that remains after the conditions of section 2 are
imposed---that of purely $x^+$-dependent gauge transformations.

\vfill\eject
\references

\refis{bandd}J.D. Bjorken and S.D. Drell, {\it Relativistic Quantum Fields}
	(McGraw-Hill, 1965)

\refis{bpv93}C. Bender, S. Pinsky, and B. van de Sande, \pr, D48, 1993, 816,

\refis{pv94}S. Pinsky and B. van de Sande, \pr, D49, 1994, 2001,

\refis{bmpp93}S.J. Brodsky, G. McCartor, H.-C. Pauli, and S. Pinsky,
	{\sl Particle World \bf 3 \rm (1993) 109}

\refis{burkardt89}M. Burkardt, \np, A504, 1989, 762,

\refis{dancoff50}S.M. Dancoff, \pr, 78, 1950, 382,

\refis{dirac49}P.A.M. Dirac, \rmp, 21, 1949, 392,

\refis{dirac64}P.A.M. Dirac, {\it Lectures on Quantum Mechanics} (Yeshiva
	University, 1964)

\refis{epb87}T. Eller, H.-C. Pauli, and S.J. Brodsky, \pr, D35, 1987, 1493,

\refis{ghpsw93} S. Glazek, A. Harindranath, S. Pinsky, J. Shigemitsu,
	and K. Wilson, \pr, D47, 1993, 1599,

\refis{gribov78}V.N. Gribov, \np, B139, 1978, 1,

\refis{hps91}A. Harindranath, R.J. Perry, and J. Shigemitsu, OSU preprint
	DOE-ER-01545-562 (1991)

\refis{hv87}A. Harindranath and J.P. Vary, \pr, D36, 1987, 1141,

\refis{hv88}A. Harindranath and J.P. Vary, \pr, D37, 1988,
	{1064, 1076},

\refis{hv88b}A. Harindranath and J.P. Vary, \pr, D37, 1988, 1076,

\refis{hksw92}T. Heinzl, S. Krusche, S. Simb\"urger, and E. Werner,
	\zphys, C56, 1992, 415,

\refis{hkw91b}T. Heinzl, S. Krusche, and E. Werner, \pl, 272B, 1991, 54,

\refis{hpb90}K. Hornbostel, H.-C. Pauli, and S.J. Brodsky,
	\pr, D41, 1990, 3814,

\refis{kp93}A.C. Kalloniatis and H.-C. Pauli, \zphys, C60, 1993, 255,

\refis{kp94}A.C. Kalloniatis and H.-C. Pauli, {\it On Zero Modes and
	Gauge Fixing in Light-Cone Quantized Gauge Theories,} Heidelberg
	preprint MPIH--V2--1994, to appear in {\sl Z. Phys. \bf C}

\refis{kpp94b}A.C. Kalloniatis, H.-C. Pauli, and S. Pinsky, {\it Dynamical
	Zero Modes and Pure Glue $QCD_{1+1}$ in Light-Cone Field Theory,}
	Heidelberg/Ohio State University preprint MPIH V9--94,
	OHSTPY--HEP--TH--94--001, submitted to {\sl Phys. Rev. \bf D}

\refis{kp92}M. Kaluza and H.-C. Pauli, \pr, D45, 1992, 2968,

\refis{kpw92}M. Krautg\"artner, H.-C. Pauli, and F. W\"olz, \pr, D45,
	1992, 3755,

\refis{my76}T. Maskawa and K. Yamawaki, \ptp, 56, 1976, 270,

\refis{mr92}G. McCartor and D.G. Robertson, \zphys, C53, 1992, 679,

\refis{mr94}G. McCartor and D.G. Robertson, {\it Light-Cone
	Quantization of Gauge Fields,} SMU preprint SMUHEP/93--20,
	to appear in {\sl Z. Phys. \bf C}

\refis{mp93}Y. Mo and R.J. Perry, \jcp, 108, 1993, 159,

\refis{pb85}H.-C. Pauli and S.J. Brodsky, \pr, D32, 1985, {1993, 2001},

\refis{dgr93}D.G. Robertson, \pr, D47, 1993, 2549,

\refis{tamm45}I. Tamm, \jps, 9, 1945, 449,

\refis{tangphd}A.C. Tang, Ph.D. thesis, Stanford University (1990)

\refis{tbp91}A.C. Tang, S.J. Brodsky, and H.-C. Pauli, \pr, D44, 1991, 1842,

\refis{wilsonunpub}K.G. Wilson, unpublished

\refis{wwhzpg94}K.G. Wilson, T.S. Walhout, A. Harindranath, W.-M. Zhang,
	R.J. Perry, and S.D. Glazek, {\it Nonperturbative QCD: A
	Weak-Coupling Treatment on the Light Front,} Ohio State
	University preprint (1994)

\refis{wittman88}R.S. Wittman, in {\it Nuclear and Particle
	Physics on the Light Cone,} M.B. Johnson and L.S. Kisslinger,
	eds. (World Scientific, 1988)

\refis{wh93}J.J. Wivoda and J.R. Hiller, \pr, D47, 1993, 4647,

\refis{gaugerefs}See for example:
	N. Manton, \anphys, 159, 1985, {220;},
	F. Palumbo, \pl, B243, 1990, {109;},
	E. Langmann and G.W. Semenoff, \pl, B303, 1993, {303;},
	J.E. Hetrick, \journal Nucl. Phys. {\bf B}
	(Proc. Suppl.), 30, 1993, {228;},
	F. Lenz, H.W.L. Naus, and M. Thies, Erlangen preprint, October 1993

\refis{kpp94}A.C. Kalloniatis, H.C. Pauli, S.S. Pinsky, work in
	preparation

\refis{krinprep}A.C. Kalloniatis and D.G. Robertson, work in
	preparation

\refis{mrunpub}G. McCartor and D.G. Robertson, unpublished

\endreferences
\endit
\end